\documentclass[conference]{IEEEtran}
\IEEEoverridecommandlockouts
\usepackage{amsmath,amssymb,amsfonts}
\usepackage{cite}

\usepackage{algorithmic}
\usepackage{stfloats}
\usepackage{graphicx}
\usepackage{textcomp}
\usepackage{xcolor}
\usepackage{color}
\usepackage{pifont}
\usepackage{amsthm}

\usepackage{hyperref}
\usepackage{mathrsfs}
\usepackage{booktabs}
\usepackage{hyperref}
\usepackage{cases}
\usepackage{comment}
\usepackage{empheq} 
\usepackage{caption}
\usepackage{subcaption}
\allowdisplaybreaks[4]

\usepackage{url}  
\usepackage{graphicx}  
\def\BibTeX{{\rm B\kern-.05em{\sc i\kern-.025em b}\kern-.08em
    T\kern-.1667em\lower.7ex\hbox{E}\kern-.125emX}}
\begin{document}

\title{Movable Antenna-Aided Federated Learning with Over-the-Air Aggregation: Joint Optimization of Positioning, Beamforming, and User Selection}

\author{Yang Zhao\textsuperscript{1},~\IEEEauthorblockN{Yue Xiu\textsuperscript{2},~\textup{Minrui Xu}$^{1}$,~\textup{Ning Wei}$^{2}$}
\IEEEauthorblockA{\textsuperscript{1}College of Computing and Data Science,
Nanyang Technological University, Singapore\\
\textsuperscript{2}University of Electronic Science and Technology of China, Chengdu, China
}
}

\maketitle

\begin{abstract}
Federated learning (FL) in wireless computing effectively utilizes communication bandwidth, yet it is vulnerable to errors during the analog aggregation process. While removing users with unfavorable channel conditions can mitigate these errors, it also reduces the available local training data for FL, which in turn hinders the convergence rate of the training process. To tackle this issue, we propose the use of movable antenna (MA) techniques to enhance the degrees of freedom within the channel space, ultimately boosting the convergence speed of FL training. Moreover, we develop a coordinated approach for uplink receiver beamforming, user selection, and MA positioning to optimize the convergence rate of wireless FL training in dynamic wireless environments. This stochastic optimization challenge is reformulated into a mixed-integer programming problem by utilizing the training loss upper bound. We then introduce a penalty dual decomposition (PDD) method to solve the mixed-integer mixed programming problem. Experimental results indicate that incorporating MA techniques significantly accelerates the training convergence of FL and greatly surpasses conventional methods.
\end{abstract}


\section{Introduction}
Federated Learning (FL) is a distributed machine learning framework that facilitates collaborative model development among multiple users, enabling each participant to contribute to a shared global model while retaining data locally\cite{b1}. A parameter server plays a crucial role by aggregating the local model updates these users provide to enhance the global model. However, in wireless FL settings, especially when many users are participating, transmitting information between users and the server can strain limited communication resources. In these scenarios, traditional orthogonal multiple access methods may not enable users to transmit model updates concurrently within the constraints of the available bandwidth.

To alleviate the communication bottleneck in wireless FL, a method utilizing analog signal aggregation has been introduced \cite{b3}. In this approach, several users can concurrently send local models on a common wireless channel through analog modulation, automatically allowing the receiver to aggregate these models via signal superposition. This technique enhances bandwidth efficiency and reduces communication latency compared to conventional orthogonal multiple access methods, garnering increasing interest \cite{b4}. Furthermore, recent research has led to the development of FL system prototypes that integrate wireless aggregation functionalities \cite{b7}.

Wireless computing systems that facilitate FL are especially vulnerable to noise, which can lead to the propagation of aggregation errors during the computation and communication cycles of FL \cite{b10}. Additionally, users experiencing poor channel conditions can significantly impair aggregation quality. This is because users with stronger signals must lower their transmission power to accommodate the weaker ones, resulting in a decreased overall signal-to-noise ratio (SNR) \cite{b11}. While filtering out users with poor channels can decrease aggregation errors, it also reduces the amount of available training data, which in turn hampers the overall learning performance. Recently, MA technology has emerged, allowing for the dynamic adjustment of antenna positions to optimize channel conditions, thereby enhancing both communication efficiency and sensing abilities \cite{b12}. MA-assisted systems greatly boost received power and minimize outage probability, outperforming traditional Fixed Position Antenna (FPA) systems in wireless communication effectiveness \cite{b13}. As a result, the implementation of MA technology presents a novel approach to tackling challenges related to channel degradation and user selection in FL training. In this context, this paper explores the use of wireless FL supported by MA technology and proposes a joint optimization strategy to improve the convergence speed of FL training \cite{b14}. Our goal is to reduce the global training loss after $T$ rounds of communication by designing uplink receiver beamforming and user selection methods\cite{b15}. In contrast to previous studies, we account for the dynamic fluctuations in channel conditions between the server and users over time, enabling real-time modifications to user selection, MA positioning, and beamforming strategies based on channel state information for each communication round.

The primary contribution of this paper is the formulation of a joint optimization problem involving receiver beamforming, MA positioning, and user selection as a finite-time stochastic optimization challenge for MA-supported FL systems utilizing over-the-air (OTA) techniques. We use the training loss upper bound in\cite{b6} and present an optimization approach based on penalty dual decomposition (PDD) to enhance convergence rates. The PDD method incorporates a greedy algorithm for user selection and applies successive convex approximation (SCA) to optimize beamforming and MA positioning. Simulation results indicate that MA-assisted FL not only accelerates convergence speed but also decreases computational complexity compared to traditional methods, particularly when a large number of users are involved.

\section{System Model and Problem Formulation}
\subsection{System Model}
\begin{figure}[h]
  \centering
  \includegraphics[scale=0.2]{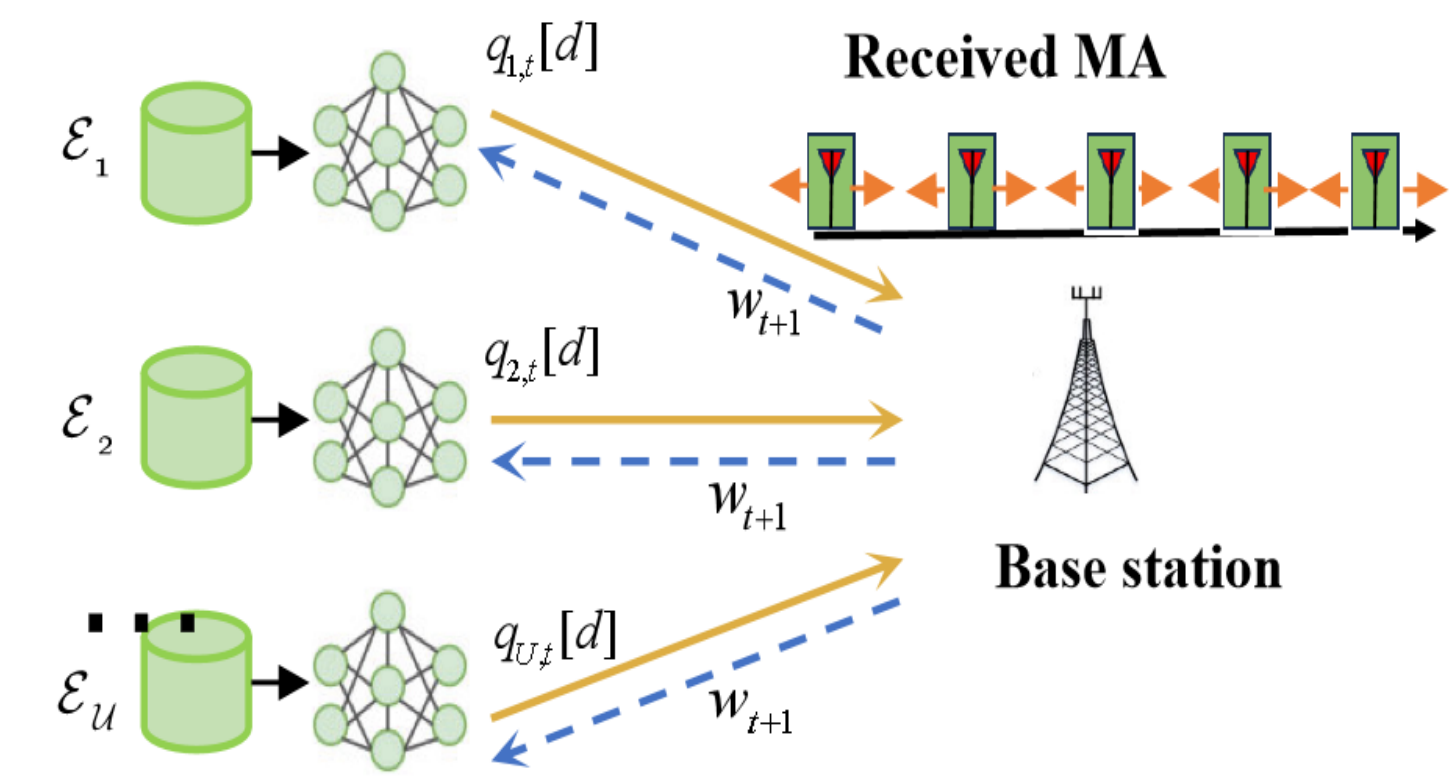}
  \captionsetup{justification=centering}
  \caption{Illustration of the MA-aided FL Systems with OTA.}
\label{FIGUREICC_0}
\end{figure}
As shown in Fig.~\ref{FIGUREICC_0}, the paper considers a wireless system consisting of a central server and $U$ local users, and the local users are represented by the set $\mathcal{U}=\{1,\ldots, U\}$. Each user 
$u$ possesses a local dataset containing 
$S_{u}$ samples, represented as $\mathcal{E}_{u}=\{(\boldsymbol{x}_{u,d},y_{u,d}):1\leq d\leq S_{u}\}$
, where 
$\boldsymbol{x}_{u,d}\in\mathbb{R}^{b}$
  denotes the feature vector of the 
$d$-th sample, and 
$y_{u,d}$
signifies the corresponding label. The users work together to train a global model on the server, which allows for precise label predictions for data points from all users while ensuring the privacy of their datasets, and the empirical loss function for the user $u$ is expressed as follows
\begin{small}
\begin{align}
F_{u}(\boldsymbol{w};\mathcal{E}_{u})=\frac{1}{S_{u}}\sum_{d=1}^{S_{u}}\mathcal{L}(\boldsymbol{w};\boldsymbol{x}_{u,d},y_{u,d}),\label{pro1}
\end{align}    
\end{small}%
where $\boldsymbol{w}\in\mathbb{R}^{D}$
represents the parameter vector of the global model, and $\mathcal{L}$ denotes the loss function computed for each individual data sample during training. The overall training loss for the global model is formulated as a weighted aggregate of the local loss functions from all participating users, expressed as $F(\boldsymbol{w})=\frac{1}{S}\sum_{u=1}^{U}S_{u}F_{u}(\boldsymbol{w};\mathcal{E}_{u}),$
where $S=\sum_{u=1}^{U}S_{u}$.
Then, we employ the Federated Stochastic Gradient Descent (FedSGD) approach for training in FL. In this method, the server refreshes the global model parameters by aggregating gradients derived from the local loss functions of chosen users. The goal is to optimize the global model $\boldsymbol{w}^{*}$ to minimize the total loss function $F(\boldsymbol{w})$. Each communication round consists of selecting a group of users, disseminating the model parameters, calculating local gradients, transmitting these gradients back to the server, and subsequently updating the global model. To account for communication noise, the server adjusts the model updates based on the real components of the received gradients, applying a learning rate modification.

The server is equipped with $1D$ linear array with
$N_{T}$ MAs, while each user has a single antenna. The uplink LoS channel between user $u$ and the server during communication round $t$ is represented by $\boldsymbol{h}_{u,t}\in\mathbb{C}^{N_{T}}$, and it is given by
$\boldsymbol{h}_{u,t}=\beta_{u,t}\boldsymbol{a}(\boldsymbol{x}_{u,t})$,
where $\beta_{u,t}$ is the complex path gain, and $\boldsymbol{a}(\boldsymbol{x}_{u,t})$=$[$ $e^{j2\pi/\lambda x_{u,t,1}\cos(\theta)}$,$e^{j2\pi/\lambda x_{u,t,2}\cos(\theta)}$,$\ldots,e^{j2\pi/\lambda x_{u,t,N_{T}}\cos(\theta)}$ $]^{T}$ $\in\mathbb{C}^{N_{T}\times 1}$, $\boldsymbol{x}_{u,t}=[x_{u,t,1}$,$x_{u,t,2}$,$\ldots$,$x_{u,t,N_{T}}]^{T}\in\mathbb{C}^{N_{T}\times 1}$ denotes the MA position, $\theta$ is the angle of arrival (AoA) of the line-of-sight
(LoS) channel. 
To efficiently combine local gradients, we employ OTA analog aggregation through a multiple access channel. As shown in Fig.~\ref{FIGUREICC_0}, during each communication round, users transmit their local gradients simultaneously on the same frequency band. This produces a combined signal at the server, which receives a weighted sum of the local gradients. In the $t$-th round, each selected user employs $D$ symbol durations to send their local gradient. The local gradient vector $\boldsymbol{g}_{u,t}$ for each user $u$ is first normalized by a scalar $v_{u,t}$, and then multiplied by a transmission weight $a_{u,t}$ prior to transmission. The transmitted signal over the $n$-th component of the local gradient, represented as $f_{u,t}[n]$, is expressed as follows
\begin{align}
f_{u,t}[n]=a_{u,t}g_{u,t}[n]/v_{u,t}.\label{pro4}
\end{align}
To support server processing, each chosen user $u$ transmits its local gradient normalization factor 
$v_{u,t}=\frac{\|\boldsymbol{g}_{u,t}\|}{\sqrt{D}}$ via the uplink signaling channel. It is assumed that this channel is a dedicated digital link with flawless signal reception. Let 
$\boldsymbol{f}_{u,t}=[f_{u,t}[1],\ldots, f_{u,t}[N]]^{T}$ represent the signal vector sent for the local gradient 
$\boldsymbol{g}_{u,t}$. From equation (\ref{pro4}), the average power used to transmit each component of $\boldsymbol{g}_{u,t}$ by the user $u$ in round $t$ is given by $\frac{\|\boldsymbol{f}_{u,t}\|^{2}}{D}=|a_{u,t}|^{2}$
. The transmission power is constrained by a maximum allowable average power $P_{a}$, such that $|a_{u,t}|^{2}\leq P_{a}$, for all 
$u$ and $t$. The received signal at the server over the 
$n$-th channel is then expressed as
\begin{small}
\begin{align}
\boldsymbol{y}_{n,t}=\sum\nolimits_{u\in\mathcal{U}_{t}^{s}}\boldsymbol{h}_{u,t}f_{u,t}[n]+\boldsymbol{s}_{n,t},\label{pro5}
\end{align}
\end{small}%
where $\boldsymbol{s}_{n,t}\sim\mathcal{CN}(\boldsymbol{0},\sigma_{n}^{2}\boldsymbol{I})$ represents the additive white Gaussian noise (AWGN) at the receiver for the 
$n$-th channel use, which is independent and identically distributed (i.i.d.) over time $t$. The server employs receive beamforming to process the incoming signal. Let $\boldsymbol{q}_{t}\in\mathbb{C}^{N}$ denote the received beamforming vector during the communication round 
$t$, subject to the constraint $\|\boldsymbol{q}_{t}\|_{2}=1$. Additionally, let $\eta_{t}\in\mathbb{R}^{+}$ represent the scaling factor for reception. The received signal is represented as $\boldsymbol{y}_{n,t}$ and can be expressed as 
$\tilde{y}_{t}[n]=\boldsymbol{q}_{t}^{H}\boldsymbol{y}_{n,t}/\sqrt{\eta_{t}}.$


\subsection{Problem Formulation}
Given the importance of learning efficiency in FL, our goal is to enhance the training convergence rate. Specifically, we aim to minimize the expected global loss function after 
$T$ communication rounds. To achieve this, we focus on jointly optimizing the selection of devices $\mathcal{U}_{t}^{s}$, the transmit weights $\{a_{u,t}\}$ of the devices, and the server-side processing, including the beamforming vector 
$\boldsymbol{q}_{t}$ and the scaling factor 
$\eta_{t}$. The optimization problem can be expressed as
\begin{small}
\begin{subequations}
\begin{align} \min_{\{\boldsymbol{x}_{u,t},\boldsymbol{q}_{t},\boldsymbol{e}_{t},\eta_{t},\{a_{u,t}\}\}_{t=0}^{T-1}}&\mathbb{E}[F(\boldsymbol{w}_{T})],\label{pro6a}\\
\text{subject to}~ 
&|a_{u,t}|^{2}\leq P_{a},&\label{pro6b}\\ 
&\|\boldsymbol{q}_{t}\|_{2}=1,&\label{pro6c}\\ 
&\eta_{t}>0,&\label{pro6d}\\ 
&\boldsymbol{e}_{t}\in\{0,1\}^{U},&\label{pro6e}\\
&|x_{u,t,n_{1}}-\boldsymbol{x}_{u,t,n_{2}}|\geq v,~n_{1}\neq n_{2},&\label{pro6f}\\
&\boldsymbol{x}_{u,t}\in\mathcal{C}. &\label{pro6g}
\end{align}\label{pro6}%
\end{subequations}
\end{small}%
In this context, $\mathbb{E}$ denotes the expectation calculated over the receiver's noise, while $\boldsymbol{e}_{t}$ is a binary vector that indicates device selection for round $t$. Specifically, if the $u$-th element $\boldsymbol{e}_{t}[u]=1$, it signifies that device $u$ has been selected to contribute to model updating for communication round $t$; conversely, $\boldsymbol{e}_{t}[u]=0$ indicates it has not been selected. It is noteworthy that $\boldsymbol{e}_{t}$ and $\mathcal{U}_{t}^{s}$ provide the same information, where $\mathcal{U}_{t}^{s}=\{u:\boldsymbol{e}_{t}[u]=1,u\in\mathcal{U}\}$. 
Based on the expression of $\boldsymbol{h}_{u,t}$, it is important to note that, unlike traditional channels with FPA, the MA channel depicted by $\boldsymbol{h}_{u,t}$ depends on the location of the received MA, which affects the channel matrix as well as the corresponding global loss function $\mathbb{E}[F(\boldsymbol{w}_{T})]$. Therefore, the MA location constraint is introduced into the problem in (\ref{pro6}). $v$ represents the minimum separation distance of antennas, and $\mathcal{C}$ refers to the movement region.

The upper bound is derived based on specific assumptions about the global loss function 
$F(\boldsymbol{w})$, commonly encountered in the stochastic optimization literature \cite{b4}. Building on these assumptions and following the derivation outlined in \cite{b5}, the expected difference between the global loss function at round 
$(t+1)$ and the optimal loss can be bounded as follows
\begin{small}
\begin{align}
&\mathbb{E}[F(\boldsymbol{w}_{t+1})-F(\boldsymbol{w}^{*})]\leq\phi_{t}\mathbb{E}[F(\boldsymbol{w}_{t})-F(\boldsymbol{w}^{*})]+\nonumber\\
&\frac{\alpha_{1}}{L}r(\boldsymbol{q}_{t},\boldsymbol{e}_{t};\mathcal{T}_{t}),\label{pro10}
\end{align}
\end{small}%
where $\phi_{t}=1-\frac{\mu}{L}(1-2\alpha_{2}r(\boldsymbol{q}_{t},\boldsymbol{e}_{t};\mathcal{T}_{t}))$, $\mathcal{T}_{t}=\{\boldsymbol{h}_{u,t}:u\in\mathcal{U}\}$, and
\begin{small}
\begin{align}
&r(\boldsymbol{q}_{t},\boldsymbol{e}_{t};\mathcal{T}_{t})=\frac{4}{U^{2}}(\sum\nolimits_{u=1}^{U}(1-e_{t}[u])S_{u})^{2}+\nonumber\\
&\frac{\sigma_{n}^{2}}{P_{a}(\sum\nolimits_{u=1}^{U}e_{t}[u]S_{u})^{2}}\max\nolimits_{1\leq u\leq U}\frac{e_{t}[u]S_{u}^{2}}{|\boldsymbol{q}_{t}^{H}}\boldsymbol{h}_{u,t}|^{2},
\end{align}\label{pro11}%
\end{small}
where $\mu\geq 0$ is strongly convex constant, and $L\geq 0$ is Lipschitz constant\cite{b4}. $\alpha_{1}\geq 0$ and $\alpha_{2}\geq 1$. Let $\boldsymbol{w}_{0}$ be the initial model parameter vector. Applying the
bound in \cite{b17} to $\mathbb{E}[F(\boldsymbol{w}_{t+1})-F(\boldsymbol{w}^{*})]$ for $t=0,\ldots,T-1$,
we have the following upper bound after $T$ communication
rounds
\begin{small}
\begin{align}
&\mathbb{E}[F(\boldsymbol{w}_{T})-F(\boldsymbol{w}^{*})]\leq(\prod_{t=0}^{T-1}\phi_{t})\mathbb{E}[F(\boldsymbol{w}_{0})-F(\boldsymbol{w}^{*})]+\frac{\alpha_{1}}{L}(\nonumber\\
&
\sum_{t=0}^{T-2}(\prod_{\tau=t+1}^{T-1}\phi_{r})r(\boldsymbol{q}_{t},\boldsymbol{e}_{t};\mathcal{T}_{t}))+r(\boldsymbol{q}_{T-1},\boldsymbol{e}_{T-1};\mathcal{T}_{T-1})).\label{pro12}
\end{align}
\end{small}%

It is important to note that minimizing 
$\mathbb{E}[F(\boldsymbol{w}_{T})]$ in (\ref{pro6a}) is equivalent to minimizing 
$\mathbb{E}[F(\boldsymbol{w}_{T})-F(\boldsymbol{w}^{*})]$. However, directly optimizing 
$\mathbb{E}[F(\boldsymbol{w}_{T})-F(\boldsymbol{w}^{*})]$ is challenging. Instead, we focus on minimizing its upper bound as shown in (\ref{pro12}). Since 
$\phi_{t}$
  is a growing function of 
$r(\boldsymbol{q}_{t},\boldsymbol{e}_{t};\mathcal{T}_{t}))$, the upper bound in (\ref{pro12}) also increases with 
$r(\boldsymbol{q}_{t},\boldsymbol{e}_{t};\mathcal{T}_{t}))$, $t = 0,\ldots,T-1$. Therefore, to reduce the upper bound, it suffices to minimize 
$r(\boldsymbol{q}_{t},\boldsymbol{e}_{t};\mathcal{T}_{t}))$ at each round with respect to 
$(\boldsymbol{q}_{t},\boldsymbol{e}_{t})$. This combined optimization problem for user selection and receiver beamforming is formulated below, where the subscript 
$t$ is omitted for simplicity
\begin{small}
\begin{subequations}
\begin{align} \min_{\mathcal{U}}&\frac{4}{U^{2}}(\sum\nolimits_{u=1}^{U}(1-e[u])S_{u})^{2}+\nonumber\\
&\frac{\sigma_{n}^{2}}{P_{a}(\sum\nolimits_{u=1}^{U}e[u]S_{u})^{2}}\max\nolimits_{1\leq u\leq U}\frac{e[u]S_{u}^{2}}{|\boldsymbol{q}^{H}\boldsymbol{h}_{u}|^{2}},\label{pro13a}\\
\text{s.t}~ 
&(\ref{pro6c}),(\ref{pro6e}),(\ref{pro6f}),(\ref{pro6g}),\label{pro13b}
\end{align}\label{pro13}%
\end{subequations}
\end{small}%
in which $\mathcal{U}=\{\boldsymbol{x}_{u,k},\boldsymbol{q},\boldsymbol{e}\}$. 
Due to the non-convex nature of the problem (\ref{pro13}), directly solving it may lead to suboptimal solutions. To address this challenge, we introduce an auxiliary variable $c$, which helps reformulate the complex non-convex optimization into a more tractable form, and we let
\begin{small}
\begin{align}
c=\max\nolimits_{1\leq u\leq U}\frac{e[u]S_{u}^{2}}{|\boldsymbol{q}^{H}\boldsymbol{h}_{u}|^{2}}.\label{pro14}
\end{align}
\end{small}%
According to (\ref{pro14}), we have $(e[u]S_{u}^{2})/|\boldsymbol{q}^{H}\boldsymbol{h}_{u}|^{2}\leq c$.
Because of the complexity of the problem and following the framework of the PDD algorithm, it is essential to introduce auxiliary variables $\bar{e}[u]$, $\tilde{e}[u]$, $\hat{e}[u]$, and we have
$e[u]=\tilde{e}[u]=\bar{e}[u]=\hat{e}[u],$ and (\ref{pro6e}) is equivalently written as
$e[u](1-e[u])=0, 0\leq e[u]\leq 1.$

Then, we further introduce the auxiliary variables $\alpha_{u}$ and $\tilde{\boldsymbol{x}}_{u,n_{1}}$, and we let $\alpha_{u}=|\boldsymbol{q}^{H}\boldsymbol{h}_{u}|^{2}$ and $\tilde{\boldsymbol{x}}_{u,n_{1}}=\boldsymbol{x}_{u,n{1}}-\boldsymbol{x}_{u,n{2}}$. Therefore, the problem is rewritten as
\begin{small}
\begin{subequations}
\begin{align} \min_{\mathcal{U}}&\frac{4}{U^{2}}(\sum\nolimits_{u=1}^{U}(1-\tilde{e}[u])S_{u})^{2}+\frac{\sigma_{n}^{2}c}{P_{a}(\sum\nolimits_{u=1}^{U}e[u]S_{u})^{2}},\label{pro18a}\\
\text{s.t.}~ 
&e[u]=\tilde{e}[u]=\hat{e}[u]=\bar{e}[u],~\forall~u,&\label{pro18b}\\
&\frac{e[u]S_{u}^{2}}{\alpha_{u}}\leq c&\label{pro18c}\\
&\alpha_{u}=|\boldsymbol{q}^{H}\boldsymbol{h}_{u}|^{2}&\label{pro18d}\\
&e[u](1-e[u])=0, 0\leq e[u]\leq 1,&\label{pro18e}\\
&\boldsymbol{x}_{u}\in\mathcal{C},&\label{pro18f}\\
&|\tilde{\boldsymbol{x}}_{u,n_{1}}|_{2}\geq v,~n_{1}\neq n_{2},&\label{pro18g}\\
&\tilde{\boldsymbol{x}}_{u,n_{1}}=\boldsymbol{x}_{u,n{1}}-\boldsymbol{x}_{u,n{2}}.&\label{pro18h}
\end{align}\label{pro18}%
\end{subequations}
\end{small}%
For the non-convex constraint in (\ref{pro18f}), we use SCA to cope with constraint (\ref{pro18f}) and $\boldsymbol{x}_{u,n_{2}}$ is given as initial MA positions. Constraint (\ref{pro18g}) can be rewritten in the following form
\begin{small}
\begin{align}
&\|\tilde{\boldsymbol{x}}_{u,n_{1}}^{(t-1)}\|_{2}-2\tilde{\boldsymbol{x}}_{u,n_{1}}^{(t-1)}\tilde{\boldsymbol{x}}_{u,n_{1}}^{T}\geq v,\label{pro_18}
\end{align}
\end{small}%
where $\tilde{\boldsymbol{x}}_{u,n_{1}}^{(t-1)}=\boldsymbol{x}_{u,n_{2}}$. 
We let $\boldsymbol{a}(\boldsymbol{x}_{u})=\boldsymbol{b}_{u}$, and $\boldsymbol{b}_{u}$ is an auxiliary variable introduced to decouple the nonlinear relationship between the antenna positions and the array response matrix. Given that all elements in the matrix response must satisfy
the constant modulus constraint, we introduce the constant modulus constraint for $|\boldsymbol{b}_{u}(i)|=1$. Then, we introduce the auxiliary variable $\gamma_{u}$, and we have
$\gamma_{u}=\boldsymbol{q}^{H}\boldsymbol{b}_{u}\beta_{u}.$\label{pro20}%

\begin{figure}[t]
  \centering
  \includegraphics[width=0.48\textwidth, height=0.23\textwidth]{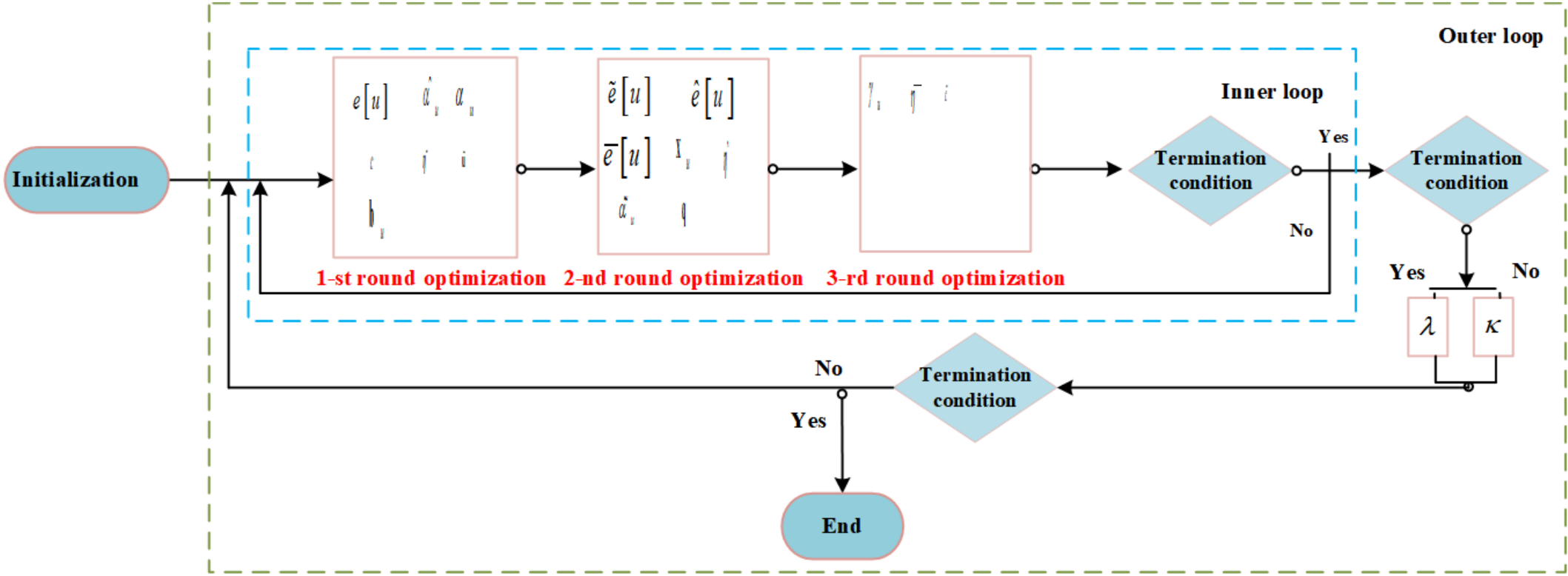}
  \captionsetup{justification=centering}
  \caption{Flowchart of the proposed algorithm..}
\label{FIGUREICC_1}
\end{figure}

Then, we continue to introduce the auxiliary variables $\eta$,$\tilde{\eta}$, $\hat{\eta}$, $\bar{\eta}$, $\tilde{\boldsymbol{u}}$, $\tilde{c}$ and let $\eta=\frac{c\sigma_{n}^{2}}{P_{a}(\sum\nolimits_{u=1}^{U}e[u]S_{u})^{2}}$, $\tilde{\eta}=\sum\nolimits_{u=1}^{U}e[u]S_{u}$,
$\hat{\eta}=\sum\nolimits_{u=1}^{U}\tilde{e}[u]S_{u}$, $\boldsymbol{u}=\tilde{\boldsymbol{u}}$, $\tilde{c}=\bar{\eta}\tilde{\eta}$. Thus, problem (\ref{pro18}) can be further expressed as
\begin{small}
\begin{subequations}
\begin{align} \min_{\bar{\mathcal{U}}}&\frac{4}{U^{2}}(U-\hat{\eta})^{2}+\eta,\label{pro21a}\\
\text{s.t.}~ 
&\bar{e}[u]S_{u}^{2}\leq\hat{\alpha}_{u},\alpha_{u}=|\gamma_{u}|^{2}, P_{a}\eta\tilde{c}= c\sigma_{n}^{2},&\label{pro21b}\\
&\tilde{\eta}=\sum\nolimits_{u=1}^{U}\hat{e}[u]S_{u},\hat{\eta}=\sum\nolimits_{u=1}^{U}\tilde{e}[u]S_{u},&\label{pro21c}\\
&|\boldsymbol{b}_{u}(i)|=1,&\label{pro21d}\\
&\tilde{\alpha}_{u}=\alpha_{u},\tilde{c}=c, \hat{\alpha}_{u}=\tilde{\alpha}_{u}\tilde{c}, \gamma_{u}=\boldsymbol{q}^{H}\boldsymbol{b}_{u}\beta_{u}, \tilde{\eta}=\hat{\eta}, &\nonumber\\
&\bar{\eta}=\hat{\eta}\tilde{\eta}, \boldsymbol{a}(\boldsymbol{x}_{u})=\boldsymbol{b}_{u}, \boldsymbol{u}=\tilde{\boldsymbol{u}},&\label{pro21e}\\
&(\ref{pro18b}),(\ref{pro18e}),(\ref{pro18f}),(\ref{pro18g}),(\ref{pro18h}),(\ref{pro_18}),&\label{pro21f}
\end{align}\label{pro21}%
\end{subequations}
\end{small}%
where $\bar{\mathcal{U}}=\{e[u],\tilde{e}[u],\bar{e}[u],\hat{e}[u],\alpha_{u},\tilde{\alpha}_{u},\hat{\alpha}_{u},\gamma_{k},\eta,\bar{\eta},\hat{\eta},\tilde{\eta},c,\tilde{c},$ $\boldsymbol{u},\tilde{\boldsymbol{u}},\boldsymbol{b}_{u},\boldsymbol{x}_{u,n_{1}},\tilde{\boldsymbol{x}}_{u,n_{1}}\}$. Following the execution framework of the PDD-based algorithm, the penalty parameter $\kappa$ is updated using the formula
$\kappa(t+1) = c\kappa(t)$, $(0<c<1)$, depending on the level of constraint violation, and the dual variables are adjusted based on (\ref{pro22}) at the top of the next page, where $t$ is the number of outer iterations. Following the approach outlined in \cite{b16}, the flow of the proposed PDD-based algorithm is summarized in Fig.~\ref{FIGUREICC_1}. The dual variables are given by
\begin{small}
\begin{align}
&\lambda_{\mathcal{U}_{1}}^{(t+1)}=\lambda_{\mathcal{U}_{1}}^{(t)}+\frac{1}{\kappa^{(t)}}\chi,\boldsymbol{\lambda}_{\boldsymbol{
\mathcal{U}}_{1}}^{(t+1)}=\boldsymbol{\lambda}_{\boldsymbol{
\mathcal{U}}_{1}}^{(t+1)}+\frac{1}{\kappa^{(t)}}\boldsymbol{\chi},\nonumber\\
&\|\boldsymbol{h}(\mathcal{U}_{1}^{(t)})\|_{\infty}=\max\{|\mathcal{U}_{1}|\},\|\boldsymbol{h}(\boldsymbol{\mathcal{U}}_{1}^{(t)})\|_{\infty}=\max\{|\boldsymbol{\mathcal{U}}_{1}|\},\label{pro_22}%
\end{align}
\end{small}%
in which $\mathcal{U}_{1}\in\{(e[u]-\tilde{e}[u]),(e[u]-\hat{e}[u]),(e[u]-\bar{e}[u]),(\gamma_{u}-\boldsymbol{q}^{H}\boldsymbol{b}_{u}\beta_{u}),(\alpha_{u}-\tilde{\alpha}_{u}),(c-\tilde{c}),(\hat{\alpha}_{u}-\tilde{\alpha}_{u}\tilde{c}),(\tilde{\eta}-\hat{\eta}),(\bar{\eta}-\hat{\eta}\tilde{\eta})\}$
and $\boldsymbol{\mathcal{U}}_{1}\in\{(\boldsymbol{a}(\boldsymbol{x}_{u})-\boldsymbol{b}_{u}),(\boldsymbol{q}-\tilde{\boldsymbol{q}})\}$.
$t$ represents the number of outer iterations. As illustrated in Fig.~\ref{FIGUREICC_1}, the proposed algorithm’s inner loop organizes the optimized variables into three blocks, which are updated sequentially. However, during each optimization round, the variables within each block can be updated in parallel. Once the inner loop meets the specified accuracy tolerance, the algorithm progresses to the outer loop, where it updates the dual variables and penalty parameters until convergence to a stationary solution set is achieved. A detailed explanation of the update procedure follows.
\vspace{-5pt}
\begin{figure*}
\begin{small}
\begin{subequations}
\begin{align} 
\min_{\bar{\mathcal{U}}}&\frac{4}{K^{2}}(M-\hat{\eta})^{2}+\eta+\frac{1}{2\kappa}((\sum\nolimits_{u=1}^{U}|e[u]-\tilde{e}[u]+\kappa\lambda_{\tilde{e}_{u}}|^{2}+|e[u]-\hat{e}[u]+\kappa\lambda_{\hat{e}_{u}}|^{2}+|e[u]-\bar{e}[u]+\kappa\lambda_{\bar{e}_{u}}|^{2}+|\gamma_{u}-\boldsymbol{q}^{H}\boldsymbol{b}_{u}\beta_{u}+\kappa\lambda_{\gamma_{u}}|^{2}\nonumber\\
&+\|\boldsymbol{a}(\boldsymbol{x}_{u})-\boldsymbol{b}_{u}+\kappa\lambda_{\boldsymbol{b}_{u}}\|^{2}+\|\alpha_{u}-\tilde{\alpha}_{u}+\kappa\lambda_{\tilde{\alpha}_{u}}\|^{2}+\|\hat{\alpha}_{u}-\tilde{\alpha}_{u}\tilde{c}+\kappa\lambda_{\hat{\alpha}_{u}}\|^{2}+\|\tilde{\boldsymbol{x}}_{u,n_{1}}-(\boldsymbol{x}_{u,n_{1}}-\boldsymbol{x}_{u,n_{2}})+\kappa\boldsymbol{\lambda}_{\tilde{\boldsymbol{x}}_{u,n_{1}}}\|^{2})\nonumber\\
&+\|c-\tilde{c}+\kappa\lambda_{\tilde{c}}\|^{2}+|\tilde{\eta}-\hat{\eta}+\kappa\lambda_{\tilde{\eta}}|^{2}+|\bar{\eta}-\hat{\eta}\tilde{\eta}+\kappa\lambda_{\bar{\eta}}|^{2}+\|\boldsymbol{q}-\tilde{\boldsymbol{q}}+\kappa\boldsymbol{\lambda}_{\boldsymbol{q}}\|^{2},\label{pro22a}\\
\text{s.t.}~ 
&(\ref{pro18e}), (\ref{pro18f}), (\ref{pro18g}), (\ref{pro18h}),,(\ref{pro_18}), (\ref{pro21b}),(\ref{pro21c}),(\ref{pro21d}). &\label{pro22b}
\end{align}\label{pro22}%
\end{subequations}
\end{small}%
\hrulefill
\end{figure*}

\subsection{1-st round of optimization}
The optimization problem of $e[u]$ is given by
\begin{small}
\begin{subequations}
\begin{align} 
\min_{e[u]}&|e[u]-\tilde{e}[u]+\kappa\lambda_{\tilde{e}_{u}}|^{2}+|e[u]-\hat{e}[u]+\kappa\lambda_{\hat{e}_{u}}|^{2}+\nonumber\\
&|e[u]-\bar{e}[u]+\kappa\lambda_{\bar{e}_{u}}|^{2},\label{pro23a}\\
\text{s.t.}~ 
&(1-e[u])e[u]=0, 0\leq e[u]\leq 1,&\label{pro23b}
\end{align}\label{pro23}%
\end{subequations}
\end{small}%
By introducing the Lagrange multiplier $\lambda_{1,u}$
to constraint (\ref{pro23b}), the optimal $e[u]$ can be expressed as
\begin{small}
\begin{align}
e[u]=\left\{\begin{matrix}
\frac{\lambda_{1,u}/2+\tilde{e}[u]+\hat{e}[u]+\bar{e}[u]-\kappa(\lambda_{\tilde{e}[u]}+\lambda_{\tilde{e}[u]}+\lambda_{\bar{e}[u]})}{3+\lambda_{1,u}}&0\leq e[u]\leq 1,\\
1&e[u]>1,\\
0&e[u]<0,
\end{matrix}\right.
\end{align}\label{pro24}
\end{small}%
and $\lambda_{1,u}$ is expressed as
\begin{small}
\begin{align}
&\lambda_{1,u}=\max\left\{0,-2(3+\tilde{e}[u]+\hat{e}[u]+\bar{e}[u]-\kappa(\lambda_{\tilde{e}[u]}+\lambda_{\hat{e}[u]}+\right.\nonumber\\
&\lambda_{\bar{e}[u]})),\left.2(\tilde{e}[u]+\hat{e}[u]+\bar{e}[u]-\kappa(\lambda_{\bar{e}[u]}+\lambda_{\tilde{e}[u]}+\lambda_{\hat{e}[u]}))\right\}.\nonumber 
\end{align}
\end{small}%
Similarly, using the Lagrange multipliers method, the optimal $\alpha_{u}$, $c$, $\hat{\alpha}_{u}$, $\tilde{\eta}$ and $\tilde{\boldsymbol{x}}_{u,n_{1}}$ can be expressed as
\begin{small}
\begin{align}
&\alpha_{u}=(2\tilde{\alpha}_{u}+2\kappa\lambda_{\tilde{\alpha}_{u}}-\lambda_{2,u})/2,c=\lambda_{3}\sigma_{n}^{2}/2+\tilde{c}-\kappa\lambda_{\tilde{c}},\nonumber\\
&\hat{\alpha}_{u}=\lambda_{4,u}/2+\tilde{\alpha}_{u}\tilde{c}-\kappa\lambda_{\hat{\alpha}_{u}}, \gamma_{u}=(\tilde{\gamma}_{u}+\boldsymbol{q}^{H}\boldsymbol{b}_{u}\beta_{u}-\kappa(\lambda_{\tilde{\gamma}_{u}}\nonumber\\
&+\lambda_{\gamma_{u}}))/4, \tilde{\eta}=\frac{\hat{\eta}+\kappa\lambda_{\hat{\eta}}+\eta(\bar{\eta}-\kappa\lambda_{\bar{\eta}})+\bar{\eta}(\bar{\eta}+\kappa\lambda_{\bar{\eta}})}{1+\eta^{2}+\bar{\eta}^{2}}, \nonumber\\
&\tilde{\boldsymbol{x}}_{u,n_{1}}=(\boldsymbol{x}_{u,n_{1}}-\boldsymbol{x}_{u,n_{2}})-\boldsymbol{\lambda}_{\tilde{\boldsymbol{x}}_{u,n_{1}}}+\kappa_{5,k}\tilde{\boldsymbol{x}}_{u,n_{1}}^{(t-1)}. 
\end{align}
\end{small}%
The Lagrange multipliers 
$\alpha_{u}$, $c$ and $\hat{\alpha}_{u}$ corresponding to the optimization problems for 
$\lambda_{2,u}$, $\lambda_{3}$, 
$\lambda_{4,u}$ and $\lambda_{5,u}$ are introduced as follows
\begin{small}
\begin{align}
&\lambda_{2,u}=\max\{0,2\tilde{\alpha}_{u}+2\kappa\lambda_{\tilde{\alpha}_{u}}-2|\tilde{\gamma}_{u}|^{2}\},\lambda_{3}=\max\{0,\frac{2}{\sigma_{n}^{2}}(\frac{P_{a}c}{\sigma_{n}^{2}}+\nonumber\\
&\kappa\lambda_{c}-c)\},\lambda_{4,u}=\max\left\{0,2(\bar{e}[u]S_{u}^{2}+\kappa\lambda_{\hat{\alpha}_{u}}-\tilde{\alpha}_{u}c)\right\},\nonumber\\
&\lambda_{5,u}=\max\left\{0, (\|\tilde{\boldsymbol{x}}_{u,n_{1}}^{(t-1)}\|_{2}-v-2(\tilde{\boldsymbol{x}}_{u,n_{1}}^{(t-1)})^{T}((\boldsymbol{x}_{u,n_{1}}-\boldsymbol{x}_{u,n_{2}})\right.\nonumber\\
&\left.-\boldsymbol{\lambda}_{\tilde{\boldsymbol{x}}_{u,n_{1}}}))/(2\tilde{\boldsymbol{x}}_{u,n_{1}}^{(t-1)})^{T}\tilde{\boldsymbol{x}}_{u,n_{1}}^{(t-1)}\right\}.\label{pro27}
\end{align}
\end{small}%
The subproblem with respect to $\tilde{\boldsymbol{q}}$ can be
written as
\begin{small}
\begin{subequations}
\begin{align} 
\min_{\tilde{\boldsymbol{q}}}&|\boldsymbol{q}-\tilde{\boldsymbol{q}}+\kappa\boldsymbol{\lambda}_{\tilde{\boldsymbol{q}}}\|^{2},\label{pro28a}\\
\text{s.t.}~ 
&\|\tilde{\boldsymbol{q}}\|^{2}=1,&\label{pro28b}
\end{align}\label{pro28}%
\end{subequations}
\end{small}%
The subproblem (\ref{pro28}) can be solved with the aid of projection\cite{b18}, with the solution given by
$\tilde{\boldsymbol{q}}=(\boldsymbol{q}+\kappa\boldsymbol{\lambda}_{\tilde{\boldsymbol{q}}})/\|\boldsymbol{q}+\kappa\boldsymbol{\lambda}_{\tilde{\boldsymbol{q}}}\|$.
The subproblem for $\boldsymbol{b}_{u}$ is given by
\begin{small}
\begin{subequations}
\begin{align} 
\min_{\boldsymbol{b}_{u}}&\|\boldsymbol{a}(\boldsymbol{x}_{u})-\boldsymbol{b}_{u}+\kappa\boldsymbol{\lambda}_{\boldsymbol{b}_{u}}\|^{2}+\|\gamma_{u}-\boldsymbol{q}^{H}\boldsymbol{b}_{u}\beta_{u}+\kappa\lambda_{\gamma_{u}}\|^{2},\label{pro30a}\\
\text{s.t.}~ 
&|\boldsymbol{b}_{u}(i)|=1.&\label{pro30b}
\end{align}\label{pro30}%
\end{subequations}
\end{small}%
Due to the separability of the unit modulus constraints, the one-iteration BCD-type algorithm described in the Appendix of \cite{b17} can be effectively utilized to solve the problem (\ref{pro30}).

\subsection{2-nd round of optimization}
For the subproblem over $\tilde{e}[u]$, $\hat{e}[u]$ and $\bar{e}[u]$, we continue to adopt the Lagrange multiplier method, and they are given by
\begin{small}
\begin{align}
&\tilde{e}[u]=e[u]+\kappa\lambda_{\tilde{e}_{u}}-\lambda_{6}S_{u}/2,\hat{e}[u]=e[u]+\kappa\lambda_{\hat{e}_{u}}-\lambda_{7}S_{u}/2,\nonumber\\
&\bar{e}[u]=e[u]+\kappa\lambda_{\bar{e}_{u}}-\lambda_{8}S_{u}^{2}/2.\label{pro_30}
\end{align}
\end{small}%
The Lagrange multipliers $\lambda_{6}$, $\lambda_{7}$ and $\lambda_{8,u}$ are expressed as
\begin{small}
\begin{align}
&\lambda_{6}=\max\{0,2[\sum\nolimits_{u=1}^{U}(e[u]+\kappa\lambda_{\tilde{e}_{u}})S_{u}-\hat{\eta}]/S_{u}^{2}\},\nonumber\\
&\lambda_{7}=\max\{0,2[\sum\nolimits_{u=1}^{U}(e[u]+\kappa\lambda_{\tilde{e}_{u}})S_{u}-\tilde{\eta}]/S_{u}^{2}\},\nonumber\\
&\lambda_{8,u}=\max\{0,2[(e[u]+\kappa\lambda_{\bar{e}_{u}}-S_{u}^{2}/\hat{\alpha}_{u})/S_{u}^{2}\}.\label{pro_32}
\end{align}
\end{small}%
Then, using the gradient method, the optimal $\tilde{\alpha}_{u}$ can be expressed as
\begin{small}
\begin{align}
&\tilde{\alpha}_{u}=(\alpha_{u}+\hat{\alpha}_{u}+\kappa(\lambda_{\tilde{\alpha}_{u}}+\lambda_{\hat{\alpha}_{u}}))/(1+\tilde{c}^{2}),\\ 
&\hat{\eta}=(8M/K^{2}-2\tilde{\eta}-2\kappa\lambda\tilde{\eta})/(8/K^{2}-2),\\
&\eta=c\sigma_{n}^{2}/(P_{a}\tilde{c}), ~\text{and}~\\
&\boldsymbol{q}=(\boldsymbol{I}+\sum_{u=1}^{U}(\boldsymbol{q}^{H}\boldsymbol{b}_{u}\beta_{u})^{H}\boldsymbol{q}^{H}\boldsymbol{b}_{u}\beta_{u})^{-1}(\tilde{\boldsymbol{q}}-\kappa\boldsymbol{\lambda}_{\boldsymbol{q}}+(\gamma_{u}+\kappa\lambda_{\gamma_{u}})\nonumber\\
&\boldsymbol{q}^{H}\boldsymbol{b}_{u}\beta_{u}).\label{pro31}
\end{align}
\end{small}%

The subproblem with respect to $\boldsymbol{x}_{k}$ is given by
\begin{small}
\begin{align}
\min_{\boldsymbol{x}_{k}}\|\angle\boldsymbol{a}(\boldsymbol{x}_{k})-\angle\boldsymbol{b}_{k}+\kappa\lambda_{\boldsymbol{b}_{k}}\|^{2}.\label{pro32}      
\end{align}
\end{small}%
Since $\boldsymbol{b}_{k}$ satisfies constant modulus-constrained, problem (\ref{pro32}) is rewritten as
\begin{small}
\begin{align}
&\mathcal{J}(\boldsymbol{x}_{u,n_{1}})=\sum\nolimits_{n_{1}=1}^{N_{u}}(|\frac{2\pi}{\lambda}(\boldsymbol{\pi})^{T}\boldsymbol{x}_{u,n_{1}}-\angle\boldsymbol{b}_{u}(n_{1})|^{2}\nonumber\\
&+\sum\nolimits_{n_{2}=1}^{N_{u}}\|\tilde{\boldsymbol{x}}_{u,n_{1}}-(\boldsymbol{x}_{u,n_{1}}-\boldsymbol{x}_{u,n_{2}})+\kappa\boldsymbol{\lambda}_{\tilde{\boldsymbol{x}}_{u,n_{1}}}\|^{2}).\label{pro53}
\end{align}
\end{small}%
Based on the first-order optimal condition, $\boldsymbol{x}_{u,n_{1}}$ is given by
\begin{small}
\begin{align}
&(\boldsymbol{I}+\frac{4\pi^{2}}{\lambda^{2}}\boldsymbol{\pi}(\boldsymbol{\pi})^{T})\boldsymbol{x}_{u,n_{1}}-(\tilde{\boldsymbol{x}}_{u,n_{1}}+\sum_{n_{2}=1}^{N_{u}}\boldsymbol{x}_{u,n_{2}}+\nonumber\\
&\kappa\boldsymbol{\lambda}_{\tilde{\boldsymbol{x}}_{u,n_{1}}})-\frac{2\pi}{\lambda}\boldsymbol{\pi}\angle\boldsymbol{b}_{u}(n_{1})=0.\label{pro54}%
\end{align}
\end{small}%
The solution of the equation (\ref{pro54}) is written as (\ref{pro56}) at the top of this page, in which $\mathcal{C}^{max}$ and $\mathcal{C}^{min}$ are the maximum mobile region and minimum mobile region of the MA position coordinate $\boldsymbol{x}_{u,n_{1}}$, respectively. Similarly, we can obtain the optimal solution for $\boldsymbol{x}_{u,n_{1}}$ using the same approach.
\begin{figure*}
\begin{small}
\begin{align}
\boldsymbol{x}_{u,n_{1}}=\left\{\begin{matrix}
(\boldsymbol{I}+\frac{4\pi^{2}}{\lambda^{2}}\boldsymbol{\pi}(\boldsymbol{\pi})^{T})^{-1}((\tilde{\boldsymbol{x}}_{u,n_{1}}+\sum_{n_{2}=1}^{N_{u}}\boldsymbol{x}_{u,n_{2}}+\kappa\boldsymbol{\lambda}_{\tilde{\boldsymbol{x}}_{u,n_{1}}})+\frac{2\pi}{\lambda}\angle\boldsymbol{b}_{u}),&~\textrm{if}~ \boldsymbol{x}_{u,n_{1}}\in\mathcal{C}\\
\mathcal{C}^{max}, &~\textrm{else}~\textrm{if}~J(\mathcal{C}^{max})<J(\mathcal{C}^{min})\\
\mathcal{C}^{min}, &~\textrm{else}\\
\end{matrix}\right.,\label{pro56}
\end{align}
\end{small}%
\hrulefill
\end{figure*}

\subsection{3-rd round of optimization}
For the subproblem over $\bar{\eta}$ and $\tilde{c}$, we continue to adopt the Lagrange multiplier method, and they are given by
\begin{small}
\begin{align}
&\bar{\eta}=\frac{\tilde{\eta}\tilde{c}-\eta\tilde{\eta}+\kappa(\tilde{\eta}\lambda_{\tilde{c}}+\lambda_{\bar{\eta}})}{1+\tilde{\eta}^{2}},\\~\text{and}~
&\tilde{c}=\frac{c+\kappa\lambda_{\tilde{c}}-\tilde{\alpha}_{u}\hat{\alpha}_{u}-\kappa\lambda_{\hat{\alpha}_{u}}+\bar{\eta}\tilde{\eta}-\kappa\lambda_{\tilde{\eta}}}{1-\tilde{\alpha}_{u}^{2}}.
\end{align}
\end{small}%

\section{Numerical Results}

We examine a scenario with $M=100$ users and $N_{u}$ antennas, where the distance $d_{u}$ between each user $u$ and the parameter server is uniformly distributed:  $d_{u}\sim U[10~m,100~m]$. The path loss is modeled using the COST Hata model, expressed as $PL[dB]=139.1+35.22\log(d_{u}[km])$. We assume that device channels remain constant throughout the training process, and the channel vector for device $u$ is generated as $\boldsymbol{h}_{u,t}=\boldsymbol{h}_{u}\sim\mathcal{CN}(\boldsymbol{0},\frac{1}{PL}\boldsymbol{I})$, $\forall~t$ for all $t$. The maximum average transmit power is set to $P_{a}=0~dBm$. Each device has an independent and identically distributed (i.i.d.) local dataset consisting of $S_{u}=270$ images from different classes of the MNIST~\cite{b8} dataset. After tuning hyperparameters, we establish a unified learning rate of $\lambda=0.05$ for the global model update, with all methods utilizing the complete local batch for gradient calculation in each communication round. A training epoch consists of one communication round since the entire local batch is employed. The receiver noise power is set to $\sigma_{n}^{2}=-20$~dBm, which takes into account both thermal noise and interference. 

\begin{figure*}[!h]
  \centering
  \begin{minipage}{0.3\textwidth}
    \centering
    \includegraphics[scale=0.37]{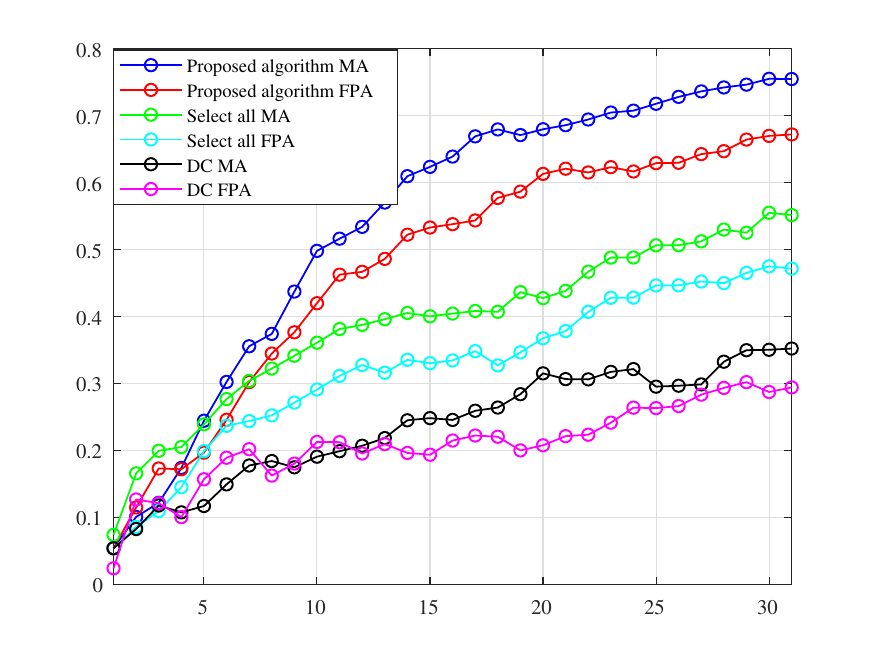}
    \captionsetup{justification=centering}
    \caption{The variation of test accuracy over epoch.}
    \label{FIGUREICC3}
  \end{minipage}
  \begin{minipage}{0.3\textwidth}
    \centering
    \includegraphics[scale=0.37]{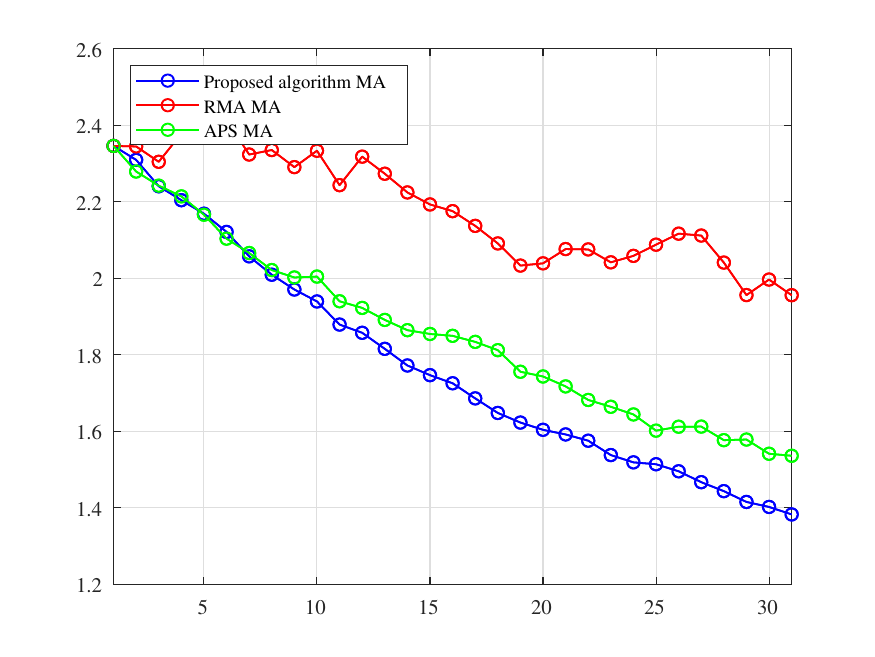}
    \captionsetup{justification=centering}
    \caption{The variation of test loss over epoch.}
    \label{FIGUREICC4}
  \end{minipage}
  \begin{minipage}{0.3\textwidth}
    \centering
    \includegraphics[scale=0.37]{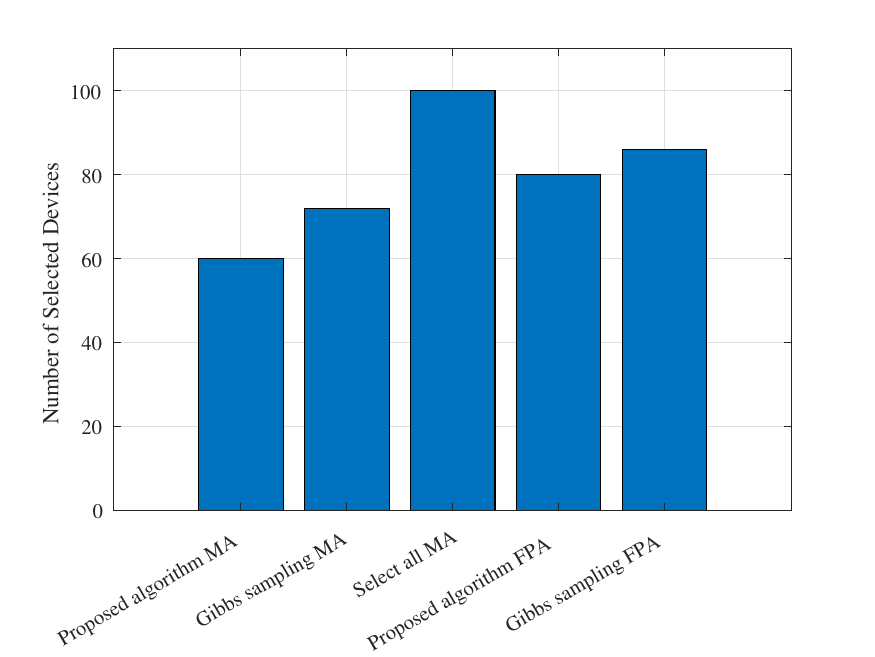}
    \captionsetup{justification=centering}
    \caption{The relationship between the number of selected devices and the MA optimization methods.}
    \label{FIGUREICC2}
  \end{minipage}
\end{figure*}
We compare our proposed method with the following two benchmark methods:
\begin{itemize}
\item 
\textbf{Select All}: All devices are selected for FL training, with beamforming and antenna positioning designed using the proposed method\cite{b7}.
\item \textbf{DC Method}: Optimize beamforming and device selection through difference of convex (DC) programming to maximize device count while keeping global model aggregation MSE within threshold $J$\cite{b7}.
\end{itemize}
Fig~\ref{FIGUREICC3} shows average test accuracy across $16$ channel settings with various noise levels. Our method converges fastest (by $T=25$ rounds) and achieves the highest accuracy, with the MA scheme exceeding 70\% test accuracy—40\% better than the DC FPA method, demonstrating the efficacy of our MA and PDD-based approach. While MA outperforms “Select All,” the latter remains more accurate than FPA-based methods, further validating MA's advantage.
Fig~\ref{FIGUREICC2} displays device selection counts across methods. The MA method selects fewer devices than FPA, and our algorithm selects fewer than others. Together with Fig~\ref{FIGUREICC3}, this confirms MA’s balance between noise resilience and training data sufficiency through local selection. For the DC method, tuning $J$ is crucial for optimal device selection. Unlike DC, our algorithm and “Select All” require no hyperparameter tuning, reaching optimal performance sooner with the same computation resources.

Fig.~\ref{FIGUREICC4} illustrates the average test loss for $16$ different channel implementations and receiver noise realizations. From the graph, we can see that our proposed PDD method outperforms the MA optimization methods based on random movable antenna (RMA) position optimization and alternating position selection (APS) presented in\cite{b13}, achieving the lowest average test loss. This is because RMA randomly designs antenna positions, and the APS scheme searches for antenna positions through quantization. However, this necessitates the use of exhaustive search algorithms, which can fail to converge and degrade performance when T is not sufficiently large.

\begin{figure}
  \centering
    \centering
    \includegraphics[scale=0.37]{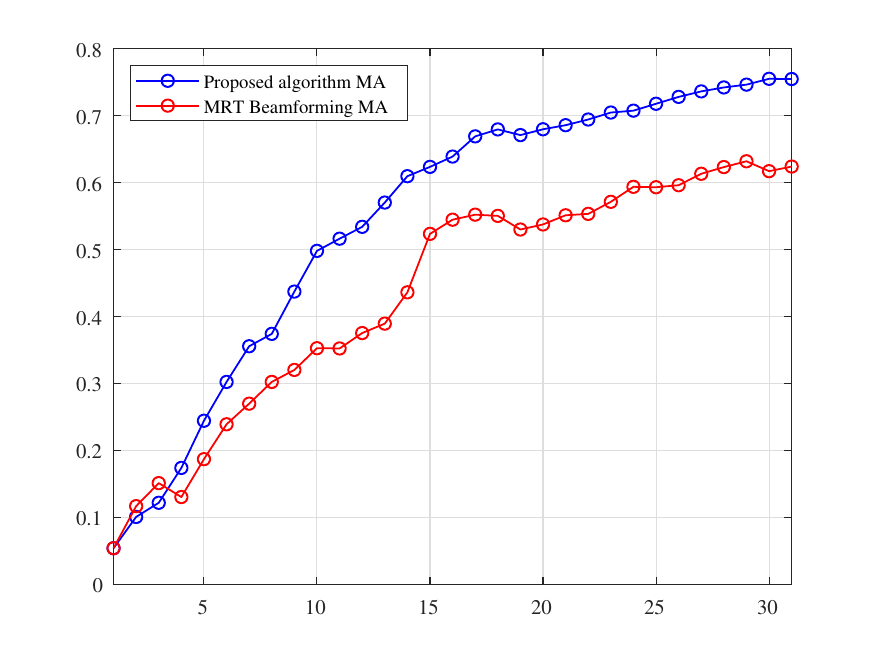}
    \captionsetup{justification=centering}
    \caption{The variation of test accuracy over epoch.}
    \label{FIGUREICC5}
\end{figure}

In Fig.~\ref{FIGUREICC5}, we compare the performance of the proposed beamforming design with the traditional maximum ratio transmission (MRT) scheme. It is evident that the proposed scheme significantly outperforms the MRT scheme in terms of testing accuracy. This is because the MRT optimizes the signal direction by maximizing signal power but does not account for the interference of other potential users. Therefore, it performs well only in single-user environments. In this paper, we consider a multi-users uplink communication scenario, which leads to the relatively poor performance of MRT.

\section{Conclusions}

In this paper, we collaboratively optimize uplink receiver beamforming, MA positioning, and device selection in MA-enabled wireless FL to minimize global training loss after any arbitrary 
$T$ communication rounds, accounting for time-varying wireless channels. To solve this non-convex problem, we use a training loss upper bound based on existing references and devise strategies for receiver beamforming, MA positioning, and device selection to minimize this bound. Our approach utilizes PDD and SCA. The PDD framework applies an alternating optimization method to iteratively address the subproblems of device selection and receiver beamforming. We also present an efficient algorithm for optimally solving the device selection subproblem with minimal computational complexity. Furthermore, simulation results illustrate that, with the chosen devices, optimizing receiver beamforming with MA significantly enhances the efficiency of the FL system compared to traditional FPA configurations.

\vspace{12pt}
\color{red}
\end{document}